# Glaucoma Detection and Structured OCT Report Generation via a Fine-tuned Multimodal Large Language Model

Jalil Jalili[1,2], Yashraj Gavhane[1,3], Evan Walker[1,2], Anna Heinke[1], Christopher Bowd[1,2], Akram Belghith[1,2], Massimo A. Fazio[4], Christopher A. Girkin[1,2], C. Gustavo De Moraes[5], Jeffrey M. Liebmann[5], Sally L. Baxter[1,2], Robert N. Weinreb[1,2], Linda M. Zangwill[1,2], and Mark Christopher*[1,2]

Author Affiliations
1. Division of Ophthalmology Informatics and Data Science, Viterbi Family Department of Ophthalmology, Shiley Eye Institute, University of California San Diego, La Jolla, CA, USA.
2. Hamilton Glaucoma Center, Viterbi Family Department of Ophthalmology, Shiley Eye Institute, University of California San Diego, La Jolla, CA, USA
3. Department of Computer Science and Engineering, University of California San Diego, La Jolla, CA, USA
4. Department of Ophthalmology and Vision Sciences, University of Alabama at Birmingham, Birmingham, AL
5. Department of Ophthalmology, Harkness Eye Institute, Bernard and Shirlee Brown Glaucoma Research Laboratory, New York, NY, United States.

Corresponding Author:
Mark Christopher, PhD,
Assistant Professor, University of California San Diego,
9500 Gilman St., San Diego, CA 92117.
mark.allen.christopher@gmail.com

Funding Sources & Disclosures
**JJ**: None
**YG:** None
**EW**: None
**AH:** None
**CB**: None
**AB**: None
**MAF:** F: National Eye Institute, EyeSight Foundation of Alabama, Research to Prevent Blindness, Heidelberg Engineering, GmbH, Topcon
**CAG:** F: National Eye Institute, EyeSight Foundation of Alabama, Research to Prevent Blindness, Heidelberg Engineering, GmbH
**CGDM**: C: Novartis; Galimedix, Belite, Reichert, Carl Zeiss, Perfuse Therapeutics; R: Heidelberg, Topcon; E: Ora Clinical
**JML: C:** Alcon, Abbvie, Carl Zeiss Meditec, Thea, Research to Prevent Blindness, R: Novartis
**SLB:** C: Topcon, F: Topcon, Optomed
**RNW:** C: Alcon,Amydis, EyeGo, Iantrek, MachineMD, Spiinogenix, Toku, Topcon; F: Centervue, Visionix, iCare, NEI, NIMHD, RPB; P: Toromedes, Zeiss Meditec
**LMZ:** C: Topcon; F: National Eye Institute, National Institutes of Health, The Glaucoma Foundation, DRCR Retina Network/JAEB Center for Health Research, The Krupp Foundation, Heidelberg Engineering; Equipment: Carl Zeiss Meditec Inc., Heidelberg Engineering GmbH, Optovue Inc., Topcon Medical Systems Inc, Icare, Optomed. P: Zeiss Meditec . Dr. Zangwill is co-founder, inventor, board member and equity holder of AISight Health Inc.
**MC:** O: AISight Health; F: NEI, The Glaucoma Foundation. Dr. Christohper is co-founder, inventor, board member and equity holder of AISight Health Inc.

Financial Support

NEI: R00EY030942, R01EY027510, R01EY034146 R01EY11008, P30EY022589, R01EY026590, EY022039, EY021818, R01EY023704, R01EY029058, EY19869, R21 EY027945, T35 EY033704, NIH: OT2OD032644. The Glaucoma Foundation. Unrestricted grant from Research to Prevent Blindness (New York, NY).

# Abstract:

**Objective:** To develop an explainable multimodal large language model (MM-LLM) that (1) screens optic nerve head (ONH) OCT circle scans for quality and (2) generates structured clinical reports that include glaucoma diagnosis and sector-wise retinal nerve fiber layer (RNFL) thinning assessments.

**Design:** Retrospective cohort study using longitudinal data from the Diagnostic Innovations in Glaucoma Study (DIGS) and the African Descent and Glaucoma Evaluation Study (ADAGES).

**Participants:** 43,849 Spectralis ONH OCT circle scans from 1,310 subjects, including 1,331 glaucomatous and 867 healthy eyes.

**Methods:** A MM-LLM (Llama 3.2 Vision-Instruct model) was fine-tuned to generate clinical descriptions of OCT imaging data. Training data included paired OCT images and automatically generated, structured clinical reports that described global and sectoral RNFL thinning. Poor-quality scans were labeled as unusable and paired with a fixed refusal statement. The model was evaluated on a held-out test set for three tasks: quality assessment, glaucoma detection, and RNFL thinning classification across seven anatomical sectors. Evaluation metrics included accuracy, sensitivity, specificity, precision, and F1-score. Model description quality was also evaluated using standard text evaluation metrics (BLEU, ROUGE, METEOR, BERTScore).

**Results:** The model achieved 0.90 accuracy and 0.98 specificity for quality triage. For glaucoma detection, accuracy was 0.86 (sensitivity 0.91, specificity 0.73, F1-score 0.91). RNFL thinning prediction accuracy ranged from 0.83 to 0.94, with highest performance in global and temporal sectors. Text generation scores (mean ± SD) showed strong alignment with reference reports (BLEU: 0.82 ± 0.19; ROUGE-1: 0.94 ± 0.08; ROUGE-2: 0.87 ± 0.17; ROUGE-L: 0.92 ± 0.11; BERTScore-F1: 0.99 ± 0.02). Stratified analysis revealed better RNFL thinning detection in moderate-to-advanced glaucoma cases, especially in temporal sectors, while performance in nasal regions was better for mild cases.

**Conclusions:** The fine-tuned MM-LLM generated accurate clinical descriptions based on OCT imaging. The model achieved high accuracy in identifying image quality issues and detecting glaucoma. The model also provided sectoral descriptions of RNFL thinning to help support clinical OCT evaluation. This approach shows potential as a scalable tool for clinical decision support, but further validation across additional datasets is needed.

## Introduction

Glaucoma is progressive optic neuropathy and a leading cause of irreversible blindness worldwide.[1] Early detection, particularly of retinal nerve fiber layer (RNFL) thinning, is critical for preserving vision, with optical coherence tomography (OCT) serving as a key imaging modality.[2] OCT-derived RNFL measurements provide essential evidence of glaucomatous structural damage, often before functional loss appears in visual field (VF) testing.[3] Although OCT enables early structural assessment, interpretation can be impeded by poor image quality and relies heavily on clinician expertise, especially when thinning patterns are subtle or complicated by comorbidities.[4] In addition, electronic health record documentation is a long-recognized burden for physicians and a known contributor to physician burnout,[5] with ophthalmologists facing particular challenges due to the high volume of patient visits and severe time constraints.[6, 7]

To address these challenges, artificial intelligence (AI) models have been proposed to assist in glaucoma detection and OCT interpretation.[8-12] While convolutional neural networks (CNNs) have demonstrated success in classification tasks, they offer limited explainability and interpretability and are generally restricted to binary or quantitative predictions.[13, 14] More recently, vision-language models (VLMs) and multimodal large language models (MM-LLMs) have emerged as promising tools for clinical applications, enabling the generation of free-text explanations based on imaging input.[15-18] Unlike traditional saliency-based methods such as Grad-CAM, MM-LLMs provide more interpretable outputs that provide a justification for model predictions.[19-23] Despite their potential, these models often suffer from hallucinations, lack of quality-awareness, and rarely offer structured, sector-wise descriptions aligned with clinical OCT reports.[24-27]

Existing approaches largely ignore the critical step of image quality assessment and do not emulate the structured format expected in ophthalmic documentation.[28-30] Moreover, few models integrate multimodal data to produce clinically grounded, interpretable outputs.[31, 32] To address these limitations, we developed a fine-tuned MM-LLM capable of (1) automatically identifying unusable OCT scans, (2) detecting glaucoma from optic nerve head (ONH) circle scans, and (3) generating concise, structured clinical reports that include sector-wise RNFL thinning assessments.

Here, we fine-tuned a MM-LLM [33, 34] using a large dataset consisting of optic nerve head (ONH) OCT imaging paired with clinical descriptions.[35] The clinical descriptions were generated based on patient glaucoma status and structural assessment. Performance was evaluated across three tasks: image quality classification, glaucoma detection, and sector-wise RNFL thinning prediction. Clinical descriptions generated by the model were also evaluated using standard text evaluation metrics (BLEU, ROUGE, METEOR, BERTScore). To our knowledge, this is the first MM-LLM designed specifically for structured ONH OCT report generation in glaucoma. Our approach includes both OCT quality assessment and glaucoma detection while providing descriptions of localized RNFL thinning. These align with common clinical OCT evaluation tasks to help provide effective, impactful clinical decision support in ophthalmology. Furthermore, LLM-generated report generation may also help alleviate some documentation burden. Our findings underscore the feasibility of automated report generation using MM-LLMs, with potential to improve both research workflows and point-of-care utility.

**Methods**

**Data Description:**

This study draws upon imaging and clinical data collected through two well-established longitudinal cohorts: the Diagnostic Innovations in Glaucoma Study (DIGS; ClinicalTrials.gov ID: NCT00221897)[36] and the African Descent and Glaucoma Evaluation Study (ADAGES; ClinicalTrials.gov ID: NCT00221923).[37] Both studies implemented harmonized, standard protocols and conducted serial ophthalmic evaluations, including optical coherence tomography (OCT) imaging and visual field (VF) testing.

A total of 43,849 Spectralis (Heidelberg Engineering, Germany) OCT circle scans, centered on the ONH, were included in this analysis. These scans were acquired from 1,310 participants, comprising 1,331 glaucomatous eyes and 867 healthy eyes, over the period from 2008 to 2021.

Glaucomatous eyes were identified based on the presence of repeatable VF defects and/or characteristic structural abnormalities of the ONH, such as neuroretinal rim thinning or localized RNFL loss, as determined by masked expert assessment of fundus photographs. Healthy eyes were required to have both normal VF results and normal optic disc appearance. Eyes showing discordant findings, such as normal fields with structural glaucomatous changes, were excluded to ensure diagnostic consistency.

VF testing was performed using the Humphrey Field Analyzer II, applying the 24-2 SITA Standard strategy. Tests exceeding established reliability thresholds, such as high fixation losses or error rates greater than 33%, were removed from consideration.

**Structured Report Generation:**

To facilitate supervised fine-tuning of the multi-modal language model, structured clinical reports were automatically generated for each OCT circle scan based on corresponding diagnostic labels and sectoral retinal nerve fiber layer (RNFL) classifications derived from the Spectralis report. These generated reports served as target text outputs during model training and were designed to emulate concise clinical documentation used in ophthalmic practice. Representative examples of these image-text training pairs are presented in Figure 1.

Each report incorporated three key components: the global glaucoma diagnosis (healthy or glaucoma), RNFL thinning status across seven anatomical sectors (global, temporal, temporal superior, temporal inferior, nasal, nasal superior, nasal inferior), and an image quality flag based on the standard UCSD Imaging Data Evaluation and Assessment (IDEA) Center assessment of the scans, indicating whether the scan was deemed usable for clinical interpretation.

Text templates were programmatically constructed to reflect the diagnostic interpretation of the scan. For example, if a scan was labeled as healthy and all RNFL sectors were within normal limits (WNL), the resulting report read: *"Based on ONH OCT image, the diagnosis is Healthy. Patient has normal RNFL thickness in all sectors."* In cases where sectoral thinning was observed, for instance, in the temporal and temporal superior regions, the description included specific mention of affected areas, such as: *"Based on ONH OCT image, the diagnosis is Glaucoma. Patient has RNFL thinning outside normal limits in the temporal and temporal superior sectors."*

For scans failing to meet the manufacturer's image quality criteria, a fixed refusal statement[38] was assigned: *"ONH OCT is unusable due to quality and/or segmentation issues."* These standardized responses prevented the model from generating potentially misleading or speculative outputs when confronted with poor quality input data.

The resulting paired dataset, composed of ONH OCT images and their automatically generated clinical descriptions, formed the foundation for supervised training.

**Model Architecture and Fine-Tuning Strategy**

This study utilized the Llama 3.2 Vision-Instruct model, an 11-billion-parameter multi-modal large language model (MM-LLM) capable of processing both text and image inputs. Llama 3.2 extends the architecture of the Llama 2 series by incorporating vision encoders and multi-modal fusion modules, allowing for image-grounded language generation. The Vision-Instruct variant is instruction-tuned to follow text prompts while attending to visual features. Its architecture combines a transformer-based image encoder with a standard decoder-only large language model, connected via a multi-layer feature projection and fusion network.[34, 35, 39]

We adopted the Unsloth implementation of the model, which supports parameter-efficient fine-tuning using LoRA (Low-Rank Adaptation)[40] and QLoRA (Quantized LoRA)[41]. Fine-tuning was performed on the "unsloth/Llama-3.2-11B-Vision-Instruct-bnb-4bit" checkpoint using 4-bit quantized weights, allowing training to be conducted efficiently on limited GPU resources (one NVIDIA A40 GPU).[42, 43] During training, the vision encoder was kept frozen, while the language layers, attention modules, and MLP layers were updated.

The model was fine-tuned on ONH OCT circle scans paired with structured clinical reports that were automatically generated based on Spectralis RNFL sector labels (within normal limits, borderline and outside normal limits) and diagnostic ground truth. Each training example consisted of a single OCT image, a standard instruction prompt ("Describe the OCT scan in detail"), and the corresponding report as the output. A low temperature value (0.1) was used during inference to reduce randomness and encourage more deterministic, clinically consistent outputs. Hyperparameters used during training, including LoRA configuration and optimization settings, are detailed in Supplementary Table S1.

**Model Evaluation:**

The fine-tuned Llama 3.2 Vision-Instruct model was evaluated on three key tasks: image quality triage, glaucoma detection, and sector-wise RNFL thinning classification. All assessments were conducted on a held-out test set (10% of all subjects) composed of ONH OCT scans excluded from model training.

For the image quality triage task, performance was measured by the model's ability to correctly identify scans that did not meet usability criteria and to generate an appropriate refusal statement. Glaucoma detection was evaluated by comparing the diagnostic impression in the generated report with the ground truth diagnosis based on standardized reading center criteria, as described earlier. Sector-wise RNFL thinning classification was assessed by matching results in the generated reports against corresponding Spectralis-derived labels in the global, temporal, temporal superior, temporal inferior, nasal, nasal superior, and nasal inferior sectors.

Classification performance across all tasks was quantified using standard metrics: accuracy, sensitivity, specificity, precision, and F1-score. A zero-rule baseline, representing the majority class, was used for comparison.

In addition to classification accuracy, the description quality of the generated structured reports was evaluated. As there is no consensus on which metric is best for evaluating AI generated text to ground truth text, we used several different metrics, each focused on a particular aspect of text comparability and each with a range between 0 and 1 (high similarity).[44-47] These metrics have been widely adopted for clinical text evaluation, including applications in radiology and ophthalmology report generation, underscoring their relevance for medical and clinical domains in addition to general natural language processing.[48-51] BLEU (Bilingual Evaluation Understudy)[44] quantifies the n-gram overlap between the generated and reference texts, with a focus on phrase-level precision. ROUGE (Recall-Oriented Understudy for Gisting Evaluation)[45] emphasizes content recall and fluency, with ROUGE-1 and ROUGE-2 evaluating unigram and bigram matches, respectively, and ROUGE-L assessing sentence-level structural similarity via the longest common subsequence. METEOR[46] accounts for synonymy and word order alignment, offering insights into semantic accuracy beyond exact lexical matches. BERTScore[47] leverages contextual embeddings from pre-trained language models to compute semantic similarity at a deeper, meaning-based level. Collectively, these metrics provide a robust, multi-dimensional assessment of how closely the model-generated reports align with expert-written clinical descriptions.

**Results**

Scans from 3746 eyes of 1,310 subjects were divided into training/validation (1,987 eyes from 1,180 subjects) and testing (211 eyes from 130 subjects) cohorts (Table 1). The diagnostic distribution was comparable across splits, with glaucomatous eyes comprising 60.3% (1199 eyes) in the training/validation set and 62.6% (132 eyes) in the testing set. The mean (SD) baseline age was 62.0 (15.0) years in the training/validation group and 59.9 (16.1) years in the testing group, with similar mean (SD) last-visit ages (65.5 (0.9) vs. 63.1 (2.9) years (p-value of 0.116), respectively). Most participants identified as White (51.5% and 47.7%) or Black/African American (40.8% and 46.2%), with smaller proportions identifying as Asian (~5.3% in both training/testing sets), American Indian, or Pacific Islander. Sex distribution was relatively balanced, with females representing 59.0% of the training/validation and 56.9% of the testing cohort. Over 88% of participants in both groups identified as non-Hispanic. Mean ocular characteristics were also similar across cohorts, including axial length (~24.2 mm), central corneal thickness (CCT) (~539–541 μm), intraocular pressure (IOP) (~14 mmHg), and visual field mean deviation (VF MD) (–5.14 vs. –5.50 dB), with no statistically significant differences observed.

The generated text descriptions demonstrated strong alignment with the reference reports across multiple evaluation metrics (Table 2, Figure 2). The model achieved an average BLEU score of 0.82 ± 0.19, reflecting high n-gram overlap. ROUGE-based evaluations further confirmed the quality of the outputs, with ROUGE-1, ROUGE-2, and ROUGE-L F-measures reaching 0.94 ± 0.08, 0.87 ± 0.17, and 0.92 ± 0.11, respectively, indicating consistency at the word-, phrase-, and sentence-levels. METEOR scored 0.92 ± 0.11, suggesting effective handling of synonyms and word order. BERTScore_F1 was exceptionally high (0.99 ± 0.02), pointing to near-perfect semantic similarity between predicted and reference descriptions. Figure 2 demonstrates that the majority of generated descriptions closely align with the reference reports after excluding poor-quality images to focus solely on usable predictions.

The model demonstrated strong performance across all classification tasks (Table 3), with class-level outcomes visualized in the confusion matrices (Figure 3). For image quality assessment, it achieved 0.90 accuracy, surpassing the zero-rule baseline of 0.85, with high specificity (0.98) but moderate sensitivity (0.44). In glaucoma diagnosis, the model reached an accuracy of 0.86 and an F1-score of 0.91, outperforming the zero-rule baseline (0.75). For sector-wise RNFL thinning prediction, accuracies ranged from 0.83 to 0.94. The model particularly excelled in the global and temporal sectors, especially the temporal superior and inferior, where it significantly exceeded the zero-rule baselines. For example, global sector accuracy was 0.84 vs. a 0.62 zero-rule baseline, temporal inferior was 0.86 vs. 0.55, and temporal superior was 0.83 vs. 0.59. These results highlight the model's effectiveness in detecting glaucomatous patterns in commonly affected regions. In contrast, while the model showed high accuracy in the nasal sectors (0.89–0.94), these values were close to or slightly below the zero-rule baselines, reflecting class imbalance and a tendency to predict "no thinning." This suggests performance in these regions is influenced more by data distribution than true discriminative ability. Figure 4 also presents qualitative examples of accepted and refused scans, as well as cases where the model's predictions either closely matched or diverged from the actual clinical descriptions.

In the stratified analysis by glaucoma severity (Table 4), diagnostic and image quality classification performance remained consistent across severity groups, with image quality accuracy at 0.90 for mild and 0.86 for moderate-to-advanced glaucoma, and glaucoma diagnosis accuracy at 0.81 in both groups. However, the model's performance in RNFL thinning prediction varied notably across retinal regions. In moderate-to-advanced glaucoma, accuracy reached 0.94 in the global sector, 0.87 in the temporal superior, and 0.97 in the temporal inferior sector, substantially higher than corresponding accuracies in mild cases (0.76, 0.77, and 0.80, respectively) with p-value < 0.0005. Conversely, in the nasal sectors, the model performed significantly better in mild glaucoma, with accuracies of 0.92 (nasal superior), and 0.94 (nasal inferior), compared to 0.71, and 0.78 in the moderate-to-advanced group (p-value < 0.0005). These findings suggest the model excels at detecting pronounced thinning in advanced glaucoma, while improvements are needed to enhance sensitivity to early-stage changes, particularly in less affected nasal regions. Based on Supplementary Table S2, no significant differences in diagnostic performance were observed across different age groups.

Supplementary Figures S1 and S2 present a comparison between the fine-tuned model and the original, non-fine-tuned Llama 3.2 model. When prompted with a general instruction, the original model often generates vague and non-specific descriptions, lacking the diagnostic precision required in clinical settings. Even when guided by the structured prompt (mirroring the format used during fine-tuning) and evaluated with a low temperature setting (0.1), the original model frequently defaults to labeling all images as "healthy," RNFL thickness as "within normal limits," and image quality as "usable." These findings underscore the importance of domain-specific fine-tuning in enabling the model to generate accurate, structured, and clinically meaningful ONH OCT reports.

**Discussion**

This study demonstrates that fine-tuned multimodal language models (MM-LLMs) can generate structured, explainable clinical reports from OCT scans with high fidelity. By delivering both accurate glaucoma detection and clinically grounded interpretability, these models represent a significant step toward the real-world integration of AI-assisted diagnostics.

By generating structured, human-like clinical reports from OCT scans, the model not only achieves high diagnostic accuracy, but also provides explanations that align closely with clinical reasoning. This reasoning-based interpretability helps bridge the gap between AI predictions and clinician judgment, potentially improving diagnostic confidence and patient care. The report generated by the model could also serve as a draft for clinician documentation, with potential to make clinical workflows more efficient and reduce documentation burden for ophthalmologists.

A key innovation in this study is the integration of an image quality triage mechanism. Poor-quality OCT scans can mislead AI models and trigger hallucinated outputs, statements that sound plausible but are clinically inaccurate.[52-54] By automatically identifying unusable scans and returning a fixed refusal statement, the model avoids producing speculative interpretations based on unreliable inputs. This safeguard is critical for clinical deployment, where erroneous outputs may misguide decision-making or erode trust in AI systems.[30] Quality triage ensures that generated reports are grounded in diagnostically valid data, helping to prevent misleading interpretations from poor-quality scans and thus supporting model transparency.

Furthermore, the use of MM-LLMs enhances explainability compared to traditional methods like Grad-CAM, which often produce coarse heatmaps without explicit rationale.[55, 56] In contrast, MM-LLMs articulate the reasoning behind predictions in natural language, improving transparency and making the outputs more actionable for clinicians. This approach not only improves interpretability but also aligns with the growing need for transparency in AI-driven diagnostics, making it easier for clinicians to trust and integrate these models into their practice.[21, 22]

We opted to freeze the vision encoder during training to reduce computational overhead and prevent the loss of general visual representations. This design choice, supported by prior multimodal architectures,[57-60] allowed the language components to adapt effectively to clinical report generation while preserving robust image embeddings.

The model consistently performed well across evaluation metrics, demonstrating its ability to generate accurate and semantically rich reports from ONH OCT scans. High BLEU and ROUGE scores suggest strong syntactic alignment with reference reports, while elevated METEOR and BERTScore values highlight the model's grasp of semantic content. Part of this strong performance likely stems from the structured nature of the target reports, enabling the model to learn consistent templates and improve similarity metrics.

Performance was especially strong in the global and temporal inferior and superior sectors, regions commonly affected by glaucoma. These results suggest the model effectively learned prevalent thinning patterns, particularly in moderate-to-advanced disease stages. In contrast, lower performance in the nasal sectors likely reflects class imbalance and fewer thinning examples in the training data. These findings emphasize the importance of curating balanced training datasets across anatomical regions and disease severity levels to enhance generalizability and ensure equitable diagnostic performance.

The comparison between the fine-tuned and original models highlights the essential role of fine-tuning in producing accurate and clinically meaningful outputs. As shown in the supplementary figures, the original model often generates vague descriptions and misclassifies low-quality scans as usable. Fine-tuning significantly improves both diagnostic precision and image quality assessment, ensuring outputs align with structured clinical standards. These findings emphasize the necessity of domain-specific fine-tuning for reliable medical applications of large language models.[61, 62]

Beyond the challenges associated with OCT-based interpretation, several additional limitations merit consideration. Training on structured clinical reports may cause the model to overfit templated phrasing, potentially reducing adaptability to varied clinical documentation styles. However, these reports may potentially offer more detail and usable information than existing clinician-generated reports, which anecdotally are often short or non-descriptive due to the strong time constraints imposed by the high volume of patient encounters which is typical in ophthalmic practice. Moreover, observed performance disparities across retinal regions and glaucoma severities highlight the model's sensitivity to imbalanced training data. To address these concerns, future research should incorporate diverse and balanced datasets, explore cross-institutional transfer learning, and include racially and ethnically representative populations. Integrating complementary modalities, such as fundus photographs, visual field tests, and longitudinal OCT data, may further enhance diagnostic accuracy and support longitudinal disease monitoring.

As MM-LLMs advance toward clinical adoption, ensuring transparency, fairness, and human oversight is critical. Embedding interpretable reasoning in AI outputs is not only a technical strength, but a clinical necessity to mitigate automation bias and uphold patient-clinician trust.

**Conclusion**

This study demonstrates the potential of fine-tuned multimodal language models to generate structured, interpretable clinical reports from OCT scans with high diagnostic accuracy. By integrating a quality triage mechanism, the model reduces misleading outputs from poor-quality scans, supporting safety and transparency. The model's sentence-level outputs explicitly localize RNFL thinning across anatomical sectors, aligning with clinical reasoning and advancing explainability in AI-generated OCT reports. These features position our approach as a scalable solution for glaucoma decision support, as well as a potential approach for reducing clinical documentation burden. Future work integrating diverse datasets and multimodal inputs will further enhance generalizability and support safe, real-world deployment.

| | Training & Validation (n = 1180 subjects; 1987 eyes) | Testing (n = 130 subjects; 211 eyes) | p-value |
|---|---|---|---|
| **Subject-Level Characteristics** | | | |
| Baseline Age, years | 62.0 (61.1, 62.8) | 59.9 (57.1, 62.7) | 0.13 |
| Last Age, years | 65.5 (64.6, 66.4) | 63.1 (60.2, 66.0) | 0.12 |
| Race | | | |
|   American Indian/ Alaska Native | 3 (0.3%) | 0 (0.0%) | 0.82 |
|   Asian | 62 (5.3%) | 7 (5.4%) | |
|   Black or African American | 482 (40.8%) | 60 (46.2%) | |
|   Native Hawaiian or Other Pacific Islander | 3 (0.3%) | 0 (0.0%) | |
|   Unknown or Not Reported | 22 (1.9%) | 1 (0.8%) | |
|   White | 608 (51.5%) | 62 (47.7%) | |
| Sex | | | |
|   Female | 696 (59.0%) | 74 (56.9%) | 0.71 |
|   Male | 484 (41.0%) | 56 (43.1%) | |
| Ethnicity | | | |
|   Hispanic | 37 (3.1%) | 5 (3.8%) | 0.61 |
|   Not Hispanic | 1042 (88.3%) | 117 (90.0%) | |
|   Unknown or Not Reported | 101 (8.6%) | 8 (6.2%) | |
| **Eye-Level Characteristics at Latest Imaging** | | | |
| Axial Length (mm) | 24.2 (24.1, 24.2) | 24.2 (24.0, 24.5) | 0.72 |
| CCT (μm) | 538.6 (536.0, 541.2) | 541.0 (533.3, 548.8) | 0.56 |
| 24-2 VF MD (dB) | -5.14 (-5.52, -4.75) | -5.50 (-6.67, -4.34) | 0.56 |
| Spherical Equivalent (D) | -0.67 (-0.81, -0.54) | -1.02 (-1.43, -0.61) | 0.12 |
| IOP (mmHg) | 14.52 (14.26, 14.78) | 14.02 (13.23, 14.81) | 0.24 |
| Diagnosis | | | |
|   Glaucomatous | 1199 (60.3%) | 132 (62.6%) | 0.96 |
|   Non-Glaucomatous | 788 (39.7%) | 79 (37.4%) | |

Table 1. *Comparison of participant and eye-level demographic and ocular characteristics between cohort splits*

| Table 2: Summary of text description evaluation metrics (BLEU, ROUGE, METEOR, and BERTScore) | | |
|---|---|---|
| **Metric** | **Value Mean (std)** | **Interpretation** |
| **BLEU Score** | 0.82 (0.19) | High n-gram overlap with the reference text, indicating strong **word- and phrase-level** similarity. |
| **ROUGE-1 F-measure** | 0.94 (0.08) | Excellent unigram recall, showing that most individual words match the reference text (**word-level**). |
| **ROUGE-2 F-measure** | 0.87 (0.17) | Strong bigram overlap, reflecting the model's ability to capture **phrase-level** coherence. |
| **ROUGE-L F-measure** | 0.92 (0.11) | High similarity in the longest common subsequence, suggesting well-preserved **sentence-level** structure. |
| **METEOR** | 0.92 (0.11) | Incorporates **synonymy and word order alignment**, indicating semantically accurate and fluent descriptions. |
| **BERTScore_F1** | 0.99 (0.02) | Extremely high **semantic similarity** based on contextual embeddings, showing alignment in meaning beyond surface-level text. |

| Feature | Accuracy | Sensitivity | Specificity | Precision | F1-Score | Zero-Rule baseline * |
|---|---|---|---|---|---|---|
| **Image Quality** | 0.90 (0.87, 0.93) | 0.44 (0.32, 0.58) | 0.98 (0.97, 0.99) | 0.82 (0.73, 0.90) | 0.58 (0.45, 0.690) | 0.85 |
| **Glaucoma Diagnosis** | 0.86 (0.81, 0.90) | 0.93 (0.88, 0.96) | 0.65 (0.55, 0.76) | 0.89 (0.82, 0.94) | 0.91 (0.86, 0.94) | 0.75 |
| **Sector-Wise RNFL Thinning Prediction:** | | | | | | |
| **Global** | 0.84 (0.80, 0.89) | 0.88 (0.79, 0.94) | 0.82 (0.75, 0.90) | 0.76 (0.66, 0.85) | 0.81 (0.74, 0.88) | 0.62 |
| **Temporal** | 0.86 (0.81, 0.90) | 0.74 (0.59, 0.86) | 0.89 (0.85, 0.93) | 0.62 (0.42, 0.77) | 0.67 (0.52, 0.79) | 0.80 |
| **Temporal Inferior** | 0.86 (0.81, 0.91) | 0.90 (0.83, 0.95) | 0.83 (0.76, 0.90) | 0.82 (0.73, 0.89) | 0.85 (0.79, 0.91) | 0.55 |
| **Temporal Superior** | 0.83 (0.79, 0.87) | 0.83 (0.77, 0.89) | 0.824 (0.77, 0.8) | 0.77 (0.69, 0.84) | 0.80 (0.74, 0.85) | 0.59 |
| **Nasal** | 0.94 (0.91, 0.96) | 0.57 (0.26, 0.80) | 0.96 (0.94, 0.98) | 0.40 (0.19, 0.56) | 0.47 (0.24, 0.62) | 0.95 |
| **Nasal Inferior** | 0.91 (0.88, 0.95) | 0.47 (0.24, 0.71) | 0.94 (0.92, 0.97) | 0.35 (0.14, 0.59) | 0.40 (0.19, 0.61) | 0.94 |
| **Nasal Superior** | 0.89 (0.85, 0.93) | 0.50 (0.32, 0.65) | 0.93 (0.89, 0.96) | 0.39 (0.21, 0.55) | 0.44 (0.27, 0.56) | 0.92 |

Table 3: Model performance evaluation across image quality, glaucoma diagnosis, and sector-wise RNFL thinning prediction

* The zero-rule baseline: Predicts the majority class. Serves as a baseline for model performance, particularly in imbalanced datasets.

| Glaucoma Severity | Mild Glaucoma (n = 62 subjects) | | | Moderate-to-Advanced Glaucoma (n = 40 subjects) | | |
|---|---|---|---|---|---|---|
| Feature | Accuracy | Sensitivity | Specificity | Accuracy | Sensitivity | Specificity |
| Image Quality | 0.90 (0.84, 0.92) | 0.30 (0.19, 0.37) | 0.99 (0.97, 0.99) | 0.86 (0.84, 0.86) | 0.51 (0.35, 0.68) | 0.97 (0.95, 0.98) |
| Glaucoma Diagnosis | 0.81 (0.74, 0.87) | 0.89 (0.83, 0.94) | 0.65 (0.55, 0.76) | 0.81 (0.72, 0.88) | 0.99 (0.97, 1.00) | 0.65 (0.55, 0.75) |
| **RNFL Thinning:** | | | | | | |
| Global | 0.76 (0.68, 0.84) | 0.80 (0.66, 0.91) | 0.74 (0.63, 0.85) | 0.94 (0.84, 0.97) | 0.97 (0.92, 0.99) | 0.56 (0.22, 0.84) |
| Temporal | 0.85 (0.79, 0.90) | 0.57 (0.40, 0.78) | 0.88 (0.82, 0.93) | 0.80 (0.69, 0.87) | 0.89 (0.78, 0.95) | 0.69 (0.55, 0.79) |
| Temporal Superior | 0.77 (0.73, 0.81) | 0.76 (0.65, 0.86) | 0.77 (0.70, 0.84) | 0.87 (0.79, 0.93) | 0.96 (0.91, 0.99) | 0.41 (0.18, 0.65) |
| Temporal Inferior | 0.80 (0.71, 0.88) | 0.86 (0.77, 0.93) | 0.76 (0.62, 0.87) | 0.97 (0.93, 0.99) | 0.99 (0.99, 1.00) | 0.71 (0.43, 0.90) |
| Nasal | 0.96 (0.94, 0.98) | 0.19 (0.00, 0.36) | 0.97 (0.95, 0.99) | 0.87 (0.81, 0.93) | 0.86 (0.70, 0.94) | 0.87 (0.79, 0.94) |
| Nasal Superior | 0.92 (0.87, 0.96) | 0.16 (0.000, 0.4) | 0.96 (0.93, 0.99) | 0.71 (0.63, 0.78) | 0.70 (0.60, 0.82) | 0.72 (0.61, 0.80) |
| Nasal Inferior | 0.94 (0.88, 0.97) | 0.22 (0.00, 0.65) | 0.97 (0.94, 0.99) | 0.78 (0.69, 0.84) | 0.63 (0.33, 0.83) | 0.80 (0.72, 0.88) |

Table 4: Stratified diagnostic accuracy (95% CI) of the MM-LLM across glaucoma severity groups (mild vs. moderate-to-advanced) for classification tasks using OCT circle scans

| Training OCT Image Samples | Training Structured Reports |
|---|---|
| **Image Quality:** Usable. **Diagnosis:** Healthy 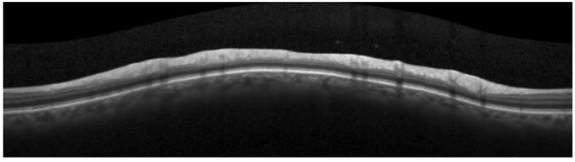 | Based on ONH OCT image, the diagnosis is Healthy. Patient has normal RNFL thickness in all sectors. |
| **Image Quality:** Usable. **Diagnosis:** Glaucoma 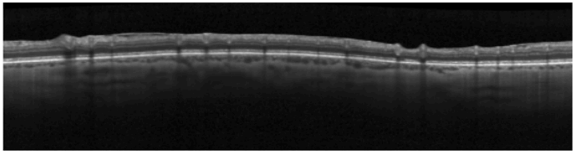 | Based on ONH OCT image, the diagnosis is Glaucoma. Patient has RNFL thinning outside normal limits in the global and temporal inferior sectors. |
| **Image Quality:** Usable. **Diagnosis:** Glaucoma 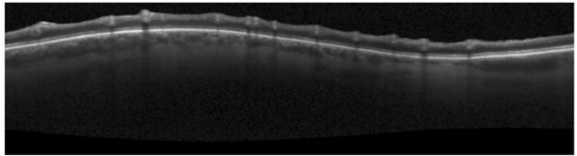 | Based on ONH OCT image, the diagnosis is Glaucoma. Patient has RNFL thinning outside normal limits in the global, temporal, temporal superior, temporal inferior, nasal, nasal superior, and nasal inferior sectors. |
| **Image Quality:** Unusable 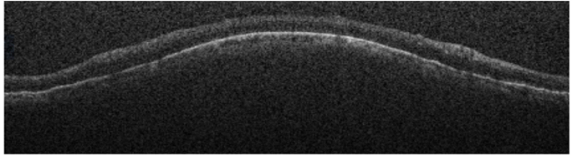 | ONH OCT is unusable due to quality and/or segmentation issues. |

Figure 1: Samples of OCT circle scan images and corresponding automatically generated structured clinical reports used for training of AI model

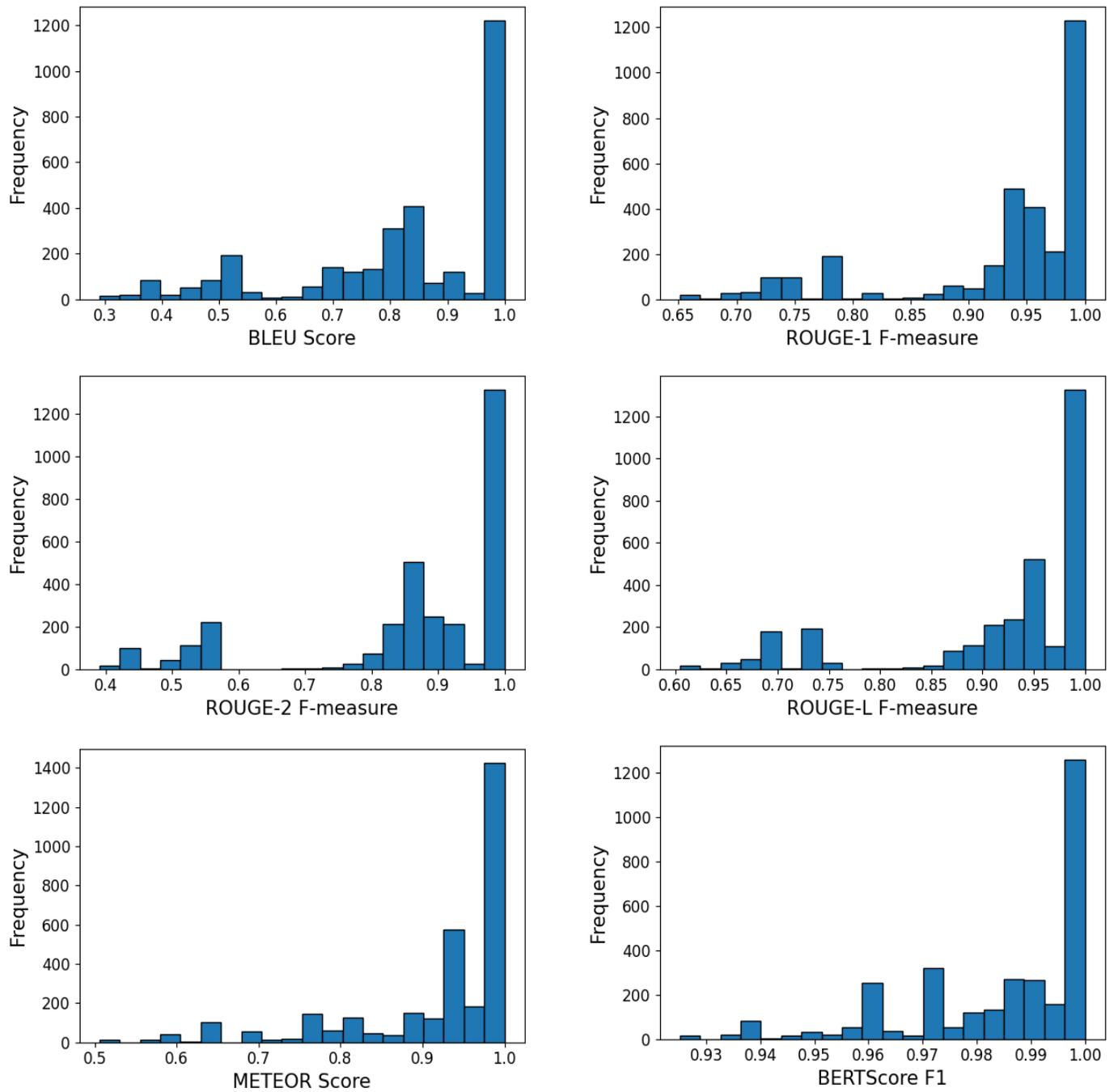

*Figure 2: Distribution histograms of BLEU, ROUGE, METEOR, and BERTScore metrics for generated text descriptions*

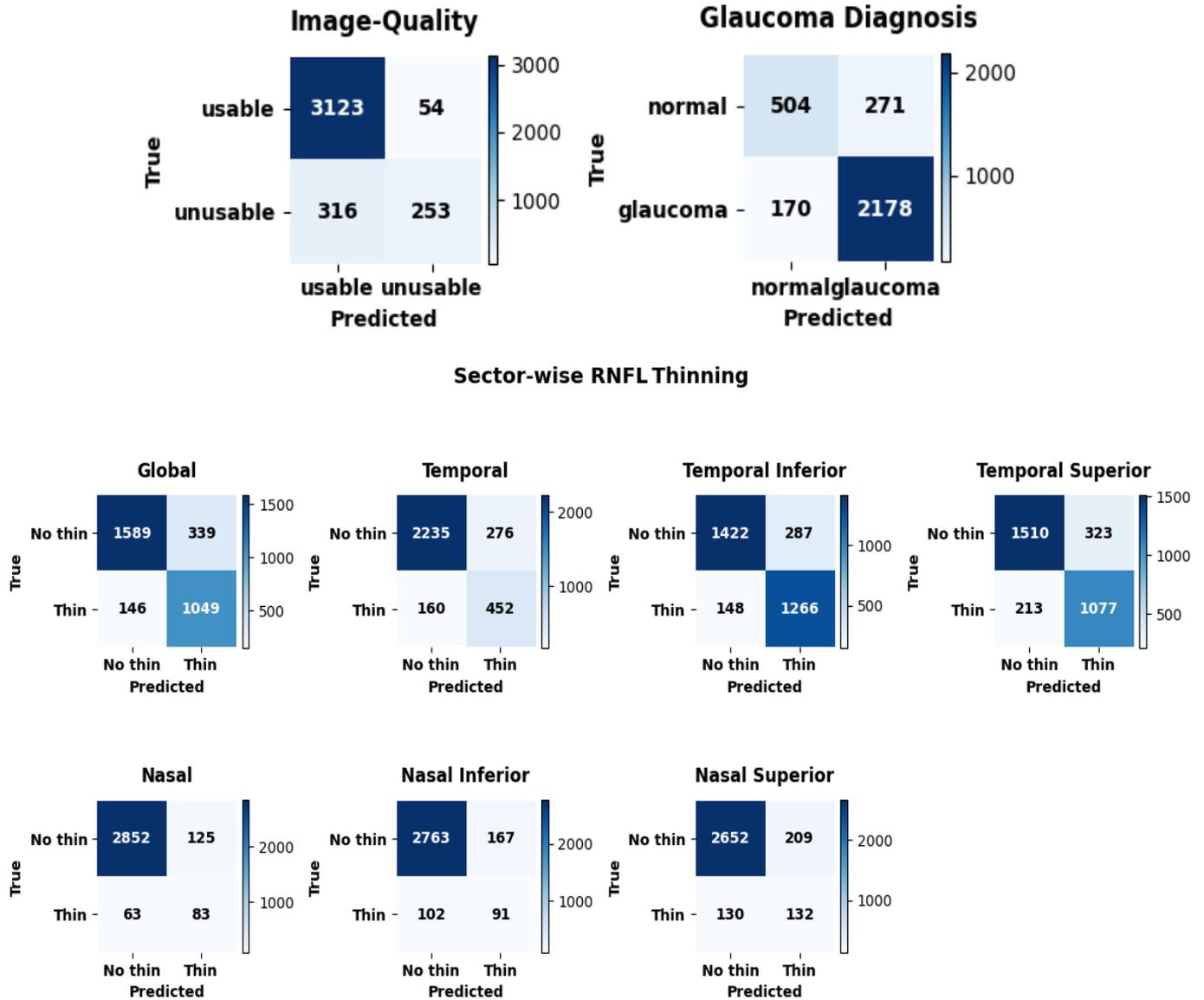

*Figure 3: Confusion matrices for image quality detection, glaucoma detection, and seven sector-wise RNFL thinning predictions*

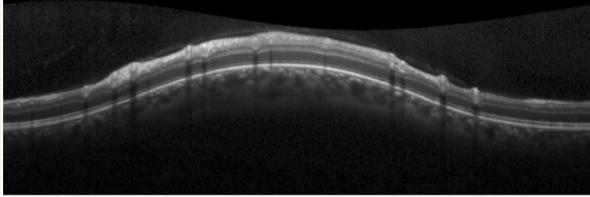

Figure 4: Examples of acceptable and unacceptable or unusable quality scans (with a refusal statement) with corresponding actual and model reports.

| Supplementary Table S1: Fine-tuning hyperparameters | |
|---|---|
| **Parameter** | Value |
| **Base model** | unsloth/Llama-3.2-11B-Vision-Instruct-bnb-4bit |
| **Precision** | 4-bit (QLoRA) |
| **LoRA rank (r)** | 16 |
| **LoRA alpha** | 16 |
| **LoRA dropout** | 0.0 |
| **Finetuned modules** | Language layers, MLP, attention |
| **Frozen modules** | Vision encoder |
| **Batch size** | 2 per device |
| **Gradient accumulation steps** | 4 |
| **Number of epochs** | 3 |
| **Learning rate** | 2e-4 |
| **Optimizer** | AdamW (8-bit) |
| **Weight decay** | 0.01 |
| **Scheduler** | Linear |
| **Max sequence length** | 2048 |
| **Instruction prompt** | "Describe the OCT scan in detail." |

| Age | Age > Median (65.3 years) (n = 65 subjects) | | | Age < Median (65.3 years) (n = 79 subjects) | | |
|---|---|---|---|---|---|---|
| Feature | Accuracy | Sensitivity | Specificity | Accuracy | Sensitivity | Specificity |
| **Image Quality** | 0.92 (0.88, 0.95) | 0.60 (0.42, 0.75) | 0.98 (0.96, 0.99) | 0.88 (0.84, 0.92) | 0.28 (0.18, 0.41) | 0.99 (0.98, 1.00) |
| **Glaucoma Diagnosis** | 0.92 (0.86, 0.96) | 0.96 (0.92, 0.99) | 0.52 (0.38, 0.66) | 0.80 (0.72, 0.87) | 0.88 (0.80, 0.94) | 0.68 (0.55, 0.81) |
| **RNFL Thinning:** | | | | | | |
| **Global** | 0.78 (0.71, 0.86) | 0.86 (0.75, 0.95) | 0.70 (0.59, 0.83) | 0.91 (0.85, 0.95) | 0.91 (0.79, 0.96) | 0.91 (0.84, 0.96) |
| **Temporal** | 0.79 (0.72, 0.86) | 0.66 (0.46, 0.86) | 0.82 (0.75, 0.89) | 0.93 (0.89, 0.96) | 0.82 (0.62, 0.92) | 0.96 (0.93, 0.98) |
| **Temporal Superior** | 0.78 (0.73, 0.82) | 0.85 (0.76, 0.92) | 0.71 (0.63, 0.79) | 0.88 (0.83, 0.92) | 0.82 (0.69, 0.92) | 0.91 (0.86, 0.95) |
| **Temporal Inferior** | 0.82 (0.74, 0.90) | 0.90 (0.81, 0.97) | 0.69 (0.58, 0.83) | 0.90 (0.85, 0.95) | 0.89 (0.77, 0.97) | 0.91 (0.86, 0.95) |
| **Nasal** | 0.94 (0.91, 0.96) | 0.39 (0.05, 0.64) | 0.96 (0.94, 0.98) | 0.94 (0.89, 0.98) | 0.67 (0.16, 0.96) | 0.96 (0.91, 0.99) |
| **Nasal Superior** | 0.87 (0.82, 0.92) | 0.40 (0.13, 0.75) | 0.90 (0.84, 0.95) | 0.91 (0.86, 0.95) | 0.56 (0.38, 0.69) | 0.96 (0.91, 0.98) |
| **Nasal Inferior** | 0.90 (0.84, 0.94) | 0.32 (0.18, 0.68) | 0.93 (0.89, 0.96) | 0.93 (0.87, 0.97) | 0.60 (0.17, 0.82) | 0.95 (0.91, 0.99) |

Supplementary Table S2: Stratified diagnostic accuracy (95% CI) of the MM-LLM across different age groups for classification tasks using OCT circle scans

**A**

General and structured prompt setups for generating outputs with the non-fine-tuned Llama 3.2 model on OCT circle scan images

◆ **General Prompt Setup**

**System Prompt:**
You are a knowledgeable assistant for medical imaging.

**User Prompt:**
Describe the OCT scan in detail!

◆ **Structured Prompt Setup**

**System Prompt:**
You are a clinical assistant trained to generate structured ONH OCT reports. Each report must follow this format:

Diagnosis: [Healthy or Glaucoma]
RNFL Assessment:

Global: [within or outside of normal limits]
Temporal: [within or outside of normal limits]
Temporal Superior: [within or outside of normal limits]
Temporal Inferior: [within or outside of normal limits]
Nasal: [within or outside of normal limits]
Nasal Superior: [within or outside of normal limits]
Nasal Inferior: [within or outside of normal limits]
Image Quality: [Usable / Unusable due to quality issues]

Generate your response based solely on the image and do not speculate beyond what is visible.

**User Prompt:**
Describe the OCT scan in detail!

**B**

Comparison between the outputs of fine-tuned and non-fine-tuned Llama 3.2 models

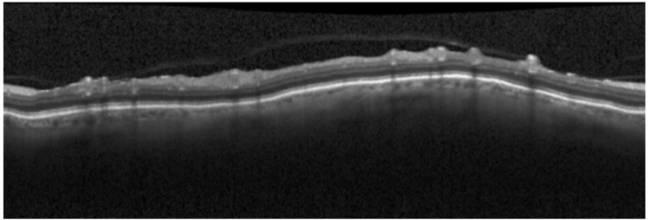

**Actual Description:** Based on ONH OCT image, the diagnosis is Glaucoma. Patient has RNFL thinning outside normal limits in the global, temporal superior, and temporal inferior sectors. **Fine-tuned Llama:** Based on ONH OCT image, the diagnosis is Glaucoma. Patient has RNFL thinning outside normal limits in the global, temporal superior, and temporal inferior sectors. **Non-fine-tuned Llama with General Prompt:** An OCT (Optical Coherence Tomography) scan is a non-invasive imaging test that uses low-coherence interferometry to capture high-resolution images of the retina and other structures in the eye. It is commonly used to diagnose and monitor various eye conditions, such as age-related macular degeneration, diabetic retinopathy, and glaucoma. **What is shown in the OCT scan?** ** The OCT scan shows a cross-sectional image of the retina, which is the light-sensitive tissue at the back of the eye. The image is composed of multiple layers, each representing different structures within the retina. **Layers of the retina:** **Retinal pigment epithelium (RPE):** ** The outermost layer of the retina, responsible for absorbing excess light and maintaining the health of the photoreceptors. * **Choroid:** ** A layer of blood vessels that supplies oxygen and nutrients to the retina. ***Sclera:** ** The white, fibrous tissue that surrounds the eye and provides protection. * **Vitreous humor:** ** The clear gel-like substance that fills the space between the lens and the retina. **Abnormalities in the OCT scan:** **Retinal thinning:** ** The retina appears thinner than normal, which can be a sign of age-related macular degeneration or other retinal diseases. * **Fluid accumulation:** ** Fluid accumulation in the retina, which can be a sign of diabetic retinopathy or other conditions. * **Scarring:** ** Scarring or atrophy of the retina, which can be a sign of previous eye injury or disease. **Interpretation of the OCT scan:** ** The OCT scan provides valuable information about the structure and health of the retina. It can help diagnose various eye conditions and monitor their progression over time. The scan can also guide treatment decisions and monitor the effectiveness of therapy. **Conclusion:** ** The OCT scan is a powerful diagnostic tool that provides detailed images of the retina and other structures in the eye. It can help diagnose and monitor various eye conditions and guide treatment decisions. **Non-fine-tuned Llama with Structured Prompt Similar to Actual Reports:** Diagnosis: Healthy **RNFL Assessment:** ** - Global: Within normal limits - Temporal: Within normal limits - Temporal Superior: Within normal limits - Temporal Inferior: Within normal limits - Nasal: Within normal limits - Nasal Superior: Within normal limits – Nasal Inferior: Within normal limits **Image Quality:

** Usable

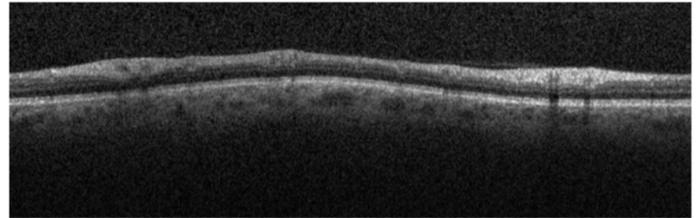

**Actual Description:** ONH OCT is unusable due to quality and/or segmentation issues.

**Fine-tuned Llama:** ONH OCT is unusable due to quality and/or segmentation issues.

**Non-fine-tuned Llama with General Prompt:** The provided OCT (Optical Coherence Tomography) scan is a cross-sectional image of the retina, showcasing the layered structure of the eye's inner surface. The scan displays a horizontal section through the retina, with the top of the image representing the surface of the retina and the bottom representing the choroid, a layer of blood vessels between the sclera and retina. The choroid is a thin, vascular layer that supplies oxygen and nutrients to the retina. The scan reveals a normal retinal structure, with distinct layers visible, including the retinal pigment epithelium (RPE), photoreceptor layer, outer nuclear layer, outer plexiform layer, inner nuclear layer, inner plexiform layer, ganglion cell layer, and nerve fiber layer. These layers are essential for transmitting visual information from the retina to the brain. The scan also shows the presence of a few small drusen, which are small yellow deposits that can accumulate under the retina and are associated with age-related macular degeneration (AMD). However, the drusen are not large enough to cause significant vision loss at this stage. Overall, the OCT scan provides valuable information about the health of the retina and can help diagnose and monitor various retinal conditions, such as AMD, diabetic retinopathy, and macular edema.

**Non-fine-tuned Llama with Structured Prompt Similar to Actual Reports:**

Diagnosis: Healthy **RNFL Assessment:** ** - Global: Within normal limits - Temporal: Within normal limits - Temporal Superior: Within normal limits - Temporal Inferior: Within normal limits - Nasal: Within normal limits - Nasal Superior: Within normal limits – Nasal

*Supplemental Figure S1: A: Prompt setups for non-fine-tuned Llama 3.2 on OCT circle scans. B: Comparison of fine-tuned vs. non-fine-tuned Llama 3.2 outputs.*

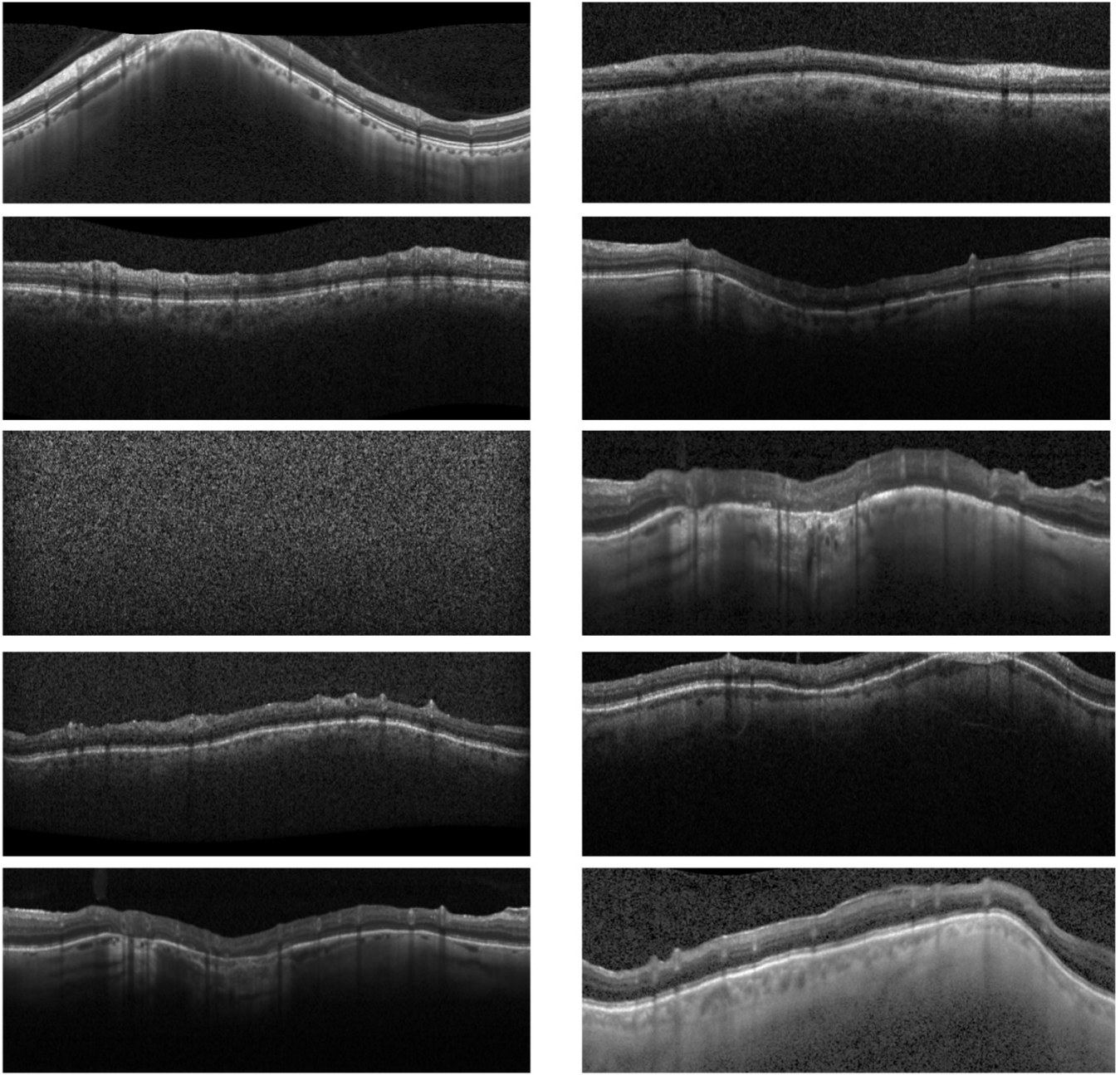

Supplemental Figure S2: Examples of poor quality scans where both ground truth and the fine-tuned model correctly identified ONH OCT scans as unusable, while the original Llama 3.2 model incorrectly labeled them as usable, underscoring the need for domain-specific instruction tuning.